\documentclass[preprint,aps,longbibliography]{revtex4-1}
\usepackage{graphicx}
\usepackage{color}
\usepackage[dvipsnames]{xcolor}
\usepackage{epstopdf, epsfig, bm}
\usepackage{hyperref}
\usepackage{amsfonts,amssymb,amsmath}

\renewcommand{\Re}{\mathrm{Re}}
\renewcommand{\Im}{\mathrm{Im}}

\begin{document}

\title{Influence of a thin compressible insoluble liquid film on the eddy currents generated by interacting surface waves}

\author{Vladimir M. Parfenyev}\email{parfenius@gmail.com}
\author{Sergey S. Vergeles}

\affiliation{
Landau Institute for Theoretical Physics, RAS, 142432, Ak. Semenova 1-A, Chernogolovka, Russia;\\
National Research University Higher School of Economics, Faculty of Physics, 101000, Myasnitskaya 20, Moscow, Russia;}

\begin{abstract}
Recently the generation of eddy currents by interacting surface waves was observed experimentally. The phenomenon provides the possibility for manipulation of particles which are immersed in the fluid. The analysis shows that the amplitude of the established eddy currents produced by stationary surface waves does not depend on the fluid viscosity in the free surface case. The currents become parametrically larger being inversely proportional to the square root of the fluid viscosity in the case when the fluid surface is covered by an almost incompressible thin liquid (i.e. shear elasticity is zero) film formed by an insoluble agent with negligible internal viscous losses as compared to the dissipation in the fluid bulk. Here we extend the theory for a thin insoluble film with zero shear elasticity and small shear and dilational viscosities on the case of an arbitrary elastic compression modulus. We find both contributions into the Lagrangian motion of passive tracers, which are the advection by the Eulerian vertical vorticity and the Stokes drift. Whereas the Stokes drift contribution preserves its value for the free surface case outside a thin viscous sublayer, the Eulerian vertical vorticity strongly depends on the fluid viscosity at high values of the film compression modulus. The Stokes drift acquires a strong dependence on the fluid viscosity inside the viscous sublayer, however, the change is compensated by an opposite change in the Eulerian  vertical vorticity. As a result, the vertical dependence of the intensity of eddy currents is given by a sum of two decaying exponents with both decrements being of the order of the wave number. The decrements are numerically different, so the Eulerian contribution becomes dominant at some depth for the surface film with any compression modulus.
\end{abstract}

\date{\today}

\maketitle

\section{Introduction}

A mass transport (relatively slow) of a fluid produced by surface waves (relatively fast) is a long-standing problem in fluid mechanics. Considering an ideal fluid, G. Stokes showed that the account of second-order corrections in the amplitude of progressive wave leads to a drift of Lagrangian particles toward the propagation direction of the wave \cite{stokes1847theory}. A control experiment should involve the generation of the monochromatic steady wave with small enough amplitude to provide a laminar regime of the transport current. It was demonstrated that the spatial distribution of the current substantially differs from the Stokes prediction, see, e.g., \cite{russell1957experimental}. The discrepancy was attributed to the Eulerian part of the mass transport. The appropriate theory was developed by M. Longuet-Higgins \cite{longuet1953mass} and it takes into account the viscosity of the fluid along with the flow nonlinearity.

The interest to the progressive wave is justified by almost unidirectional wave spectrum excited by a wind in the ocean. In the laboratory, surface waves can be excited by the wavemakers and this promoted expansion of the interest onto crossed monochromatic plane waves as the simplest nontrivial example \cite{filatov2016nonlinear,filatov2016generation,francois2017wave}. In the experiments, the eddy laminar horizontal currents induced by crossed waves were observed by analysing the velocity of floating particles by PIV-method. The eddy currents can be described in terms of vertical vorticity and the theoretical treatment of the problem \cite{filatov2016nonlinear,parfenyev2016effects} reveals both Eulerian and Stokes contributions into these currents as well. The viscosity of a fluid is shown to play a crucial role in the generation of Eulerian motion. It violates a potential approximation in the description of surface waves and leads to the appearance of a non-zero horizontal vorticity concentrated in the thin viscous sublayer in the linear approximation with respect to the wave steepness. The nonlinear interaction forces the surface tilt to produce a tilt of the horizontal vorticity that is the vertical vorticity. The generated vertical vorticity spreads by the viscous diffusion into the fluid bulk, where the Stokes drift contribution is determined solely by the potential part of the wave flow. Both these effects have comparable amplitudes and the same spatial structure near the fluid surface. Note, that the account of the corrections stemming from the vortical part of the velocity field both for Stokes drift and Eulerian current reveals that {the intensity of eddy currents is almost constant inside the viscous sublayer}.



Interestingly, the resulting expression for the velocity of floating passive tracers is independent of the fluid viscosity for the fluid with a free surface. The result can be compared to the phenomenon of the \textit{acoustic streaming} \cite[\S 80]{landau1987fluid}, where the velocity is also independent of the viscosity, even though it originates from the viscosity. The prediction is in agreement with recent experimental results \cite{francois2017wave}, where it was found that changes in the fluid viscosity by an order of magnitude do not qualitatively change horizontal drifts of floating particles driven by the surface waves. The situation should be different when the fluid surface is covered by an almost incompressible thin liquid film formed by an insoluble agent. In this case we have predicted that the velocity of floating passive tracers parametrically increases and the parameter is inversely proportional to the square root of the fluid viscosity \cite{parfenyev2016effects}. The prediction is still awaiting experimental tests.


In the present paper, we theoretically investigate the laminar eddy currents and address some new questions. The first question is how does the drift velocity of the floating passive tracers depend on the properties of the film, which possibly covers the fluid surface? Here we consider only thin liquid films and expand the model in comparison with previous works \cite{filatov2016nonlinear,parfenyev2016effects} on films with an arbitrary elastic compression modulus.
This extension allows us to develop a unified theory that includes both limiting cases and reveals the role of elastic properties of the surface film in the problem.

The fact that a surface film can modify the hydrodynamic motion is well-known since ancient times (e.g. Greeks used oil to calm the rough seas). However, the problem has no universality since the caused changes in the fluid flow substantially depend on the properties of the film itself. For example, a thin viscoelastic film of adsorbed soluble protein with a finite shear elastic modulus totally suppresses the horizontal motion of floating particles \cite{francois2015inhibition}. On the contrary, the particle motion was observed in freely suspended thin smectic film (can be considered as two-dimensional liquid), which undergoes fast transverse oscillations \cite{yablonskii2017acoustic}. The theory predicts that an almost incompressible thin liquid film formed by an insoluble agent on a fluid surface should increase the intensity of horizontal motion \cite{parfenyev2016effects}. Presumably, these differences are caused by differences in the shear properties of the films, but further detailed theoretical and experimental studies are required.

In this paper, we consider a thin (e.g. monomolecular) film which is formed by an insoluble agent, and for this reason, the film mass is conserved. Such film does not change the fluid density $\rho$ and its kinematic viscosity $\nu$. The rheological properties of the film can be characterized by four coefficients: dilational elasticity, dilational viscosity, shear elasticity and shear viscosity \cite{langevin2014rheology,lucassen1970properties}. We assume that the film is liquid, i.e. it does not resist to the shear deformations in the film plane and therefore the shear elasticity is absent. One can also neglect the dilational and shear viscosities of the film for typical experimental conditions. The approximation is justified when $\rho \nu \gg \eta_s k$, and here $\eta_s$ stands for the dilational/shear viscosity of the film and $k$ is the characteristic wave number of the flow, see, e.g., \cite{parfenyev2016nonlinear}. {In particular, this means that the dissipation due to internal viscosity of the film is small as compared to the dissipation in the fluid bulk.} In this way, the film properties can be described solely by the dilational elasticity or the compression modulus $-n (\partial \sigma/ \partial n)$, which is a positive real number under the made assumptions. Here $n$ is the film surface density and $\sigma(n)$ is the surface tension coefficient. Due to the wave motion, the periodic contractions and expansions of the fluid surface cause periodic deviations of the film density $n$ from its equilibrium value at the resting surface. The surface tension $\sigma(n)$, which depends on the surface coverage, also vary from point to point, giving rise to an additional tangential surface stress. This stress should be taken into account in the boundary condition for the fluid motion, and this is the reason why the surface film alters the wave motion. Note that when we neglect the dilational viscosity of the film we assume that there is no time lag between changes in $n$ and $\sigma$. Otherwise, the compression modulus $-n (\partial \sigma/ \partial n)$ becomes a complex number.

One of the challenges is that the film properties may be completely unknown a-priory. For this reason, we first analyse how thin compressible liquid film modifies the wave motion in the linear approximation with respect to the wave steepness and how one can infer the film properties based on the experimental observations. Then we proceed to the second-order approximation and establish the intensity and the spatial structure of eddy currents (the slow nonlinear motion of passive tracers).
We expect that the obtained results can find applications as a particle manipulation technique, see also \cite{francois2017wave,yablonskii2017acoustic} and in geophysics, since the uppermost layer of the ocean is sometimes covered by a monomolecular surface film formed by surfactant \cite{soloviev2014near, cox1954measurement}. Note here that the artificial slicks laid by boats are about a thousand times the thickness of a monomolecular layer. However, if the thickness of these slicks is the minimum length scale in the system, they can be treated as monomolecular films at least in the linear approximation \cite{jenkins1997wave}.

{The second question which we would like to address in this article is how do the generated eddy currents penetrate into the fluid bulk? The question is interesting by itself from a fundamental point of view and it is also important for applications. Moreover, most of the theoretical predictions were obtained for passive tracers, while floating particles used in experiments may demonstrate the different behaviour due to capillary effects \cite{falkovich2005surface}. Sometimes, this circumstance make it difficult to directly compare the theoretical predictions and experimental results, which are obtained by analysing the surface motion of floating particles. We believe that the measurements in the fluid bulk will provide more accurate experimental data because the capillary forces do not affect the particles which are completely immersed in the fluid.}


Our paper is structured as follows. In the next Sec.~\ref{sec:2} we introduce main ingredients of our model, define physical
quantities of interest, and write general equations for the quantities. In Sec.~\ref{sec:3} we perform a linear analysis of general equations and study how the fluid viscosity and the surface film affect the wave motion. In Sec.~\ref{sec:4} we discuss the nonlinear generation of the vertical vorticity and obtain an explicit expression for it in terms of the wave amplitudes. Then in Sec.~\ref{sec:5} we investigate the transport of a passive scalar advected by the generated vertical vorticity, taking into account the Stokes drift contribution to its motion. In Sec.~\ref{sec:6} we discuss the obtained results on particular examples and put them in the context of recent experimental studies. Finally, in Sec.~\ref{sec:7} we present a summary of our findings. Appendices contain some details of calculations.

\section{General relations}\label{sec:2}

We consider the bulk motion of an incompressible fluid of infinite depth covered by a thin liquid film, which is formed by an insoluble agent. The motion is described by the Navier-Stokes equation, see, e.g., \cite{lamb1945hydrodynamics,landau1987fluid}
\begin{equation}\label{eq:Navier-Stokes}
  \partial_t \bm v + (\bm v \cdot \nabla ) \bm v = - \nabla P/\rho + \nu \nabla^2 \bm v,
\end{equation}
where $\bm v$ is the fluid velocity, $\rho$ is the fluid mass density, $\nu$ is the kinematic viscosity coefficient, and $P$ is the modified pressure which includes the gravitational term. The equation has to be supplemented by the incompressibility condition, $\mathrm{div} \, \bm v = 0$, and by the boundary conditions posed at the fluid surface. The film on the fluid surface partially changes the boundary conditions in comparison with the free surface case, and thus affects the fluid motion.

The kinematic boundary condition remains unchanged in the presence of the film. This condition means that the fluid surface moves with the fluid velocity $\bm v$ (hereinafter the Greek indices run over $x$ and $y$, and we sum over the repeated indices)
\begin{equation}\label{eq:kinematic_BC}
  \partial_t h = v_z - v_{\alpha} \partial_{\alpha} h,
\end{equation}
where we assume that the axis $Z$ is directed opposite to the gravitational acceleration $\bm g$, the fluid surface is determined by the equation $z = h(t,x,y)$ and it coincides with the plane $z=0$ at rest. Function $h(t,x,y)$ is assumed to be single-valued that excludes wave-breaking processes from the consideration.

The film modifies the dynamic boundary condition that expresses the requirement of zero total external force acting on an arbitrary element of the fluid surface. This condition is just the second Newton law and we neglect the acceleration term because the surface film is assumed to be thin and therefore it has a negligible mass. However, one has to take into account the inhomogeneity of the surface tension coefficient $\sigma(n)$ related to its dependence on the film surface density $n$. This inhomogeneity leads to an additional tangential force and for this reason the boundary condition is modified in comparison with the free surface case. Projecting it onto the normal vector $\bm l$ to the fluid surface and on the tangential plane, we obtain
 \begin{eqnarray}
 \label{eq:DynBC_n}
 &P - 2 \rho \nu l_i l_k \partial_i v_k
 =\rho g h + \sigma K,&  \\
 &\rho \nu \delta_{ij}^{\perp} l_k (\partial_j v_k + \partial_k v_j)
 = \sigma' \delta_{ij}^{\perp}\partial_j n,&
 \label{eq:DynBC_t}
 \end{eqnarray}
see, e.g., \cite[\S 63]{landau1987fluid} or \cite{lebedev2008nearly}. Here $\delta_{ij}^{\perp} = \delta_{ij} - l_i l_j$ is the projector operator on the fluid surface, and then $\delta_{ij}^{\perp}\partial_j$ does not contain the derivative in the direction normal to the film. $K$ is the mean curvature of the surface and it is equal to the trace of the curvature tensor $K_{ik} = K_{ki} = \delta_{ij}^{\perp} \partial_j l_k$, and $\sigma' \equiv (\partial \sigma/\partial n)$. Note that one can express the unit vector normal to the fluid surface in terms of the surface elevation, $\bm l (t,x,y) =  (-\partial_x h, -\partial_y h, 1)/\sqrt{1 + (\nabla h)^2}$.

In the present study we assume that there is no time lag between the imposed film stretching and the resulting changes in surface tension, thus the surface tension coefficient $\sigma(n)$ and the surface compression modulus $-n \sigma'(n)$ depend only on the current value of the film surface density $n$. Note that we also assume that the temperature in our model is uniform. The assumption can be broken in real systems and then the model parameters like the fluid viscosity $\nu$ and the surface tension $\sigma$ will depend on the temperature distribution. This complication of the model is beyond the scope of the present paper, although it is of great interest for geophysical applications.

The set of equations becomes closed after the mass conservation law for the surface film density $n$ is added
\begin{equation}\label{eq:mass_conservation}
\partial_t n + \delta_{ij}^{\perp} \partial_{j} ( n v_i) = 0,
\end{equation}
where the value of the velocity field should be taken at the fluid surface. The equation follows from assumed insolubility of the agent which forms the film. 

In this paper, we are interested in the vortex motion, which is generated near the fluid surface due to the interaction of surface waves. It is convenient to describe such a motion in terms of the vorticity $\bm \varpi = \mathrm{curl} \, \bm v$. Taking the curl from the Navier-Stokes equation (\ref{eq:Navier-Stokes}), one obtains \cite{lamb1945hydrodynamics,landau1987fluid}
\begin{equation}\label{eq:vorticity_eq}
 \partial_t \bm \varpi
 =-(\bm v \cdot \nabla) \bm \varpi
 +(\bm\varpi \cdot \nabla) \bm v
 +\nu \nabla^2 \bm \varpi.
\end{equation}
The vorticity should also satisfy the boundary condition, which follows from equation~(\ref{eq:DynBC_t}). Acting by an operator $\epsilon_{imq} l_m \partial_q$ (it contains only the derivatives along the surface) on both sides of this equation, we find
 \begin{equation}
 l_m l_k \partial_k \varpi_m
 +(\partial_i v_k+\partial_k v_i)
 \epsilon_{imq} l_m K_{kq} = 0,
 \label{boundary2}
 \end{equation}
where $\epsilon_{ijk}$ is the unit antisymmetric tensor and $\varpi_i = \epsilon_{ijk} \partial_j v_k$. In deriving, we have used the identity $\epsilon_{imq} l_m \partial_q l_i = \epsilon_{imq} l_m K_{qi} = 0$, which can be checked by straightforward calculation. Note that the gradient of the surface tension drops from the boundary condition (\ref{boundary2}), and further this circumstance substantially simplifies the nonlinear analysis.

\section{Linear analysis}\label{sec:3}

The linear analysis of surface waves in the considered system is well-known \cite{levich1962physicochemical,lucassen1970properties}, and it shows that there are two branches, which are gravity-capillary (transverse) waves and Marangoni (longitudinal) waves. Further, we assume that the gravity-capillary waves are weakly decaying. As we will see below, this means that the dimensionless parameter $\gamma = \sqrt{\nu k^2/\omega} \ll 1$, where $\omega$ is the wave frequency and $k$ is its wave number. Marangoni waves are damped out much more rapidly, and for this reason, they are beyond our consideration. Usually, in that regime the gravity-capillary waves are described in terms of the potential approximation. In this section, we discuss how the fluid viscosity violates the approximation and how the surface film alter the amplitude of the vortical component of the flow. The details of the analysis can be found in Appendix~\ref{sec:appA}, and here we concentrate on the results.

In the linear approximation all quantities characterising the gravity-capillary waves can be expressed in terms of the surface elevation $h(t,x,y)\propto \exp(i k_{\alpha} r_{\alpha} - i \omega t)$, which is assumed to obey the small steepness condition $|\nabla h| \ll 1$. The velocity field is given by expression
\begin{equation}\label{eq:linear_velocity}
v_{\alpha} = \frac{\left( e^{{k}z} - {D} e^{{\varkappa} z} \right)}{{k} \left(1 - \frac{ k}{ \varkappa} {D} \right)} \partial_{\alpha} \partial_t h, \quad
v_z = \frac{\left( e^{{k}z} - \frac{ k}{ \varkappa} {D} e^{{\varkappa} z} \right)}{\left(1 - \frac{ k}{ \varkappa} {D} \right)} \partial_t h,
\end{equation}
where $\varkappa = \sqrt{k^2 - i \omega/\nu}$ (with a positive real part) and factor $D$ is the only parameter which depends on the film properties
\begin{equation}\label{DC}
D = \frac{2i\gamma - \varepsilon}{i\gamma \frac{\varkappa^2 + k^2}{\varkappa k}-\varepsilon},
\quad
\varepsilon = \frac{-n_0 \sigma'(n_0)}{\rho \sqrt{\nu \omega^3}/k^2},
\end{equation}
$n_0$ is an equilibrium value of the film surface density at the resting surface and $\varepsilon$ can be called dimensionless elastic stretching modulus of the film. Let us remind that the made assumption of the instant response of the film stresses on its density means that the value of surface compression elastic modulus $-n_0 \sigma'(n_0)$ is real. Besides, it should be positive in order the film to be thermodynamically stable \cite{lucassen1968longitudinal}.

Now let us discuss the obtained expression (\ref{eq:linear_velocity}). The velocity field is given by a sum of two terms. The first term is proportional to $e^{kz}$ and it corresponds to a potential part of the velocity field, which penetrates into the fluid bulk to a depth of $1/k$. The second term is proportional to $e^{\varkappa z}$ and it originates from the fluid viscosity. The penetration depth of the second term {$1/\Re \, \varkappa \sim \gamma/k$} is much smaller than $1/k$, and therefore it is localized in a thin viscous sublayer near the fluid surface. Below the viscous sublayer, on the depth of $|z| \gg \gamma/k$, one can neglect the contribution proportional to $e^{\varkappa z}$ in expression (\ref{eq:linear_velocity}) and then the velocity field becomes potential. Note that the assumption of infinite depth of fluid is correct when the depth of fluid is much larger than the penetration depth $1/k$ of the potential component.

The compressibility of the surface film comes into play through the real parameter $\varepsilon$ and the complex parameter $D$, which are related to each other, see expression (\ref{DC}). The limiting case of the film with infinite compression modulus corresponds to $\varepsilon \rightarrow \infty$, $D \rightarrow 1$ and then the film can be treated as incompressible in the linear approximation, i.e. $n(t,x,y) = n_0$. Below we refer to this situation as to the case of the almost incompressible film. If there is no film on the fluid surface then $\varepsilon \rightarrow 0$, $D \rightarrow 2 \varkappa k / (\varkappa^2 + k^2)$ and one can check that in both limiting cases expression (\ref{eq:linear_velocity}) reproduces the previously obtained results \cite{filatov2016nonlinear,parfenyev2016effects}. In the range between, the absolute value of the parameter $D$ is of the order or less than unity. Therefore, the term $kD/\varkappa$ is small at least as $\gamma \ll 1$, and this means that the surface film mainly changes the non-potential part of the horizontal velocity. Moreover, if $|D| \sim 1$ then this non-potential correction becomes comparable with the potential contribution inside the viscous sublayer. If there is no film on the fluid surface then $|D| \sim \gamma$ and the potential contribution is everywhere leading.

Next, let us calculate the vorticity $\bm \varpi = \mathrm{curl} \, \bm v$. It is produced only by the non-potential part of the velocity field and thus it should be localized in the viscous sublayer as well. The direct calculation leads to the answer
\begin{equation}\label{eq:linear_vorticity}
\varpi_{\alpha} = \epsilon_{\alpha \beta} \frac{(\varkappa^2 - k^2) D}{\varkappa k ( 1 - \frac{k}{\varkappa} D)} e^{\varkappa z} \partial_{\beta} \partial_t h, \quad \varpi_z = 0,
\end{equation}
where $\epsilon_{\alpha \beta}$ is the unit antisymmetric tensor. So, the vorticity is directed horizontally in the linear approximation. To find the vertical component we should go beyond the approximation. Note that the horizontal vorticity $\varpi_\alpha$ essentially depends on the film compressibility, since it is proportional to the parameter $D$.

The presence of the film on the fluid surface does not change the dispersion law of the surface waves, $\omega^2 = gk + \sigma_0 k^3/\rho$, except for the possible change in an equilibrium value $\sigma_0$ of the surface tension coefficient. However, the wave damping is modified
\begin{equation}\label{eq:wave_damping}
 \frac{\Im \ \omega}{\omega}
 =- \frac{\gamma}{2 \sqrt 2} \left( \Re D + \Im D \right)
 -
 \gamma^2
 +{\mathcal O}(D\gamma^2).
\end{equation}
In the case of free surface $D=\gamma \sqrt{2} (1+i) + {\mathcal O}(\gamma^2)$ and therefore first two terms in expression (\ref{eq:wave_damping}) are of the same order, and then we find $\Im \ \omega /\omega = -2 \gamma^2$. If the fluid surface is covered by the almost incompressible film, then $D=1$ and we obtain $\Im \ \omega /\omega = -\gamma/2 \sqrt{2}$. In the latter case, it would be wrong to keep the correction arising from the second term of expression (\ref{eq:wave_damping}) because the terms of the same order were omitted in deriving this expression. To conclude, the presence of a film parametrically increases the wave damping.

One can think that the influence of thin film on the velocity field and the wave damping is the strongest when the surface film is almost incompressible. However, this statement is wrong. In particular, the surface film of finite elasticity gives rise to stronger damping. We will demonstrate this on a particular example in Sec.~\ref{sec:6}. Physically, the increase of damping is caused by the resonance between the gravity-capillary and Marangoni waves.  When the gravity-capillary wave propagates, it causes local expansions and contractions of the surface film, which in turn lead to gradients of the surface tension. The motion corresponds to that in Marangoni wave and it is the most effectively excited in the resonance, where the dispersion curves of these two modes intersect. We refer the readers to the papers \cite{lucassen1968longitudinal} and \cite{alpers1989damping} for details.

\begin{table}
  \begin{center}
  \begin{tabular}{c|c|c|c}
  \hline \hline
          & $\varepsilon$   &   $D$ &  $\Im \ \omega /\omega$ \\ \hline
       almost incompressible film & $\infty$ & $1$ & $-\gamma/2 \sqrt{2}$ \\
       compressible film & $\varepsilon \gg \sqrt{\gamma}$ & $\Re D + \Im D \gg \gamma $ & Eqs. (\ref{eq:wave_damping}) and (\ref{eq:wave_damping2}) \\
       almost free surface & $\varepsilon \ll \sqrt{\gamma}$ & $\Re D + \Im D \sim \gamma$ & Eqs. (\ref{eq:wave_damping}) and (\ref{eq:wave_damping2}) \\
       free surface & $0$ & $2 \varkappa k / (\varkappa^2 + k^2)$ & $-2\gamma^2$ \\
       \hline \hline
  \end{tabular}
  \caption{The relation between film properties, values of parameters $\varepsilon$ and $D$, and the wave damping coefficient.}
  \label{tab:films}
  \end{center}
\end{table}

Note that the parameters $\varepsilon$ and $D$, which depend on the compression modulus of the film, also depend on the frequency $\omega$ and the wave number $k$ of the excited surface wave, and therefore the surface film differently affects the waves of different frequencies. At moderate values of film elastic stretching modulus {$\varepsilon\ll1$, equation (\ref{DC}) gives
\begin{equation}\label{Dsmall}
    D = (i-1)\varepsilon/\sqrt{2} + \gamma \sqrt{2} (1+i) + i\varepsilon^2 + 2 \gamma \varepsilon + {\mathcal O}(\varepsilon^3, \gamma^3, \gamma \varepsilon^2, \varepsilon \gamma^2),
\end{equation}
and thus the combination $\Re D + \Im D \approx 2\sqrt{2} \gamma + \varepsilon^2 + 2 \gamma \varepsilon$ in expression (\ref{eq:wave_damping}).} The film considerably increases the wave damping if the combination is large as compared to $\gamma$, i.e. when $\varepsilon \gg \sqrt{\gamma}$. Suppose, for example, that the film compression modulus $-n_0 \sigma'(n_0)$ does not depend on scale. Then the ratio $\varepsilon/\sqrt{\gamma}$ is proportional to $k^{7/8}$ for gravity waves and to $k^{-3/8}$ for capillary waves. Thus, the maximum of the ratio is near the point $k=k_\ast$ corresponding to the transition from gravitational to capillary waves. If the maximum value $\left( \varepsilon/\sqrt{\gamma} \right)_{k=k_\ast}\gg 1$, then there is a range of wave numbers $k_{min}<k<k_{max}$ around the point $k_\ast$ determined by inequality $\varepsilon/\sqrt{\gamma} \gtrsim 1$, where the film significantly alters the wave damping. For example, the ratio $\left( \varepsilon/\sqrt{\gamma} \right)_{k=k_\ast}$ is about $15\div35$ for some types of monomolecular films at water surface having compression elastic modulus $-n_0 \sigma'(n_0) \sim 20\div40\, \text{erg}/\text{cm}^2$ \cite{langevin2014rheology}. Outside the range the wave damping is modified only slightly. Table~\ref{tab:films} summarizes the results for the discussed limiting cases and illustrates the relation between film properties and values of parameters $\varepsilon$ and $D$.

The derived expressions can be easily generalized to the case when the surface elevation $h(t,x,y)$ is an arbitrary superposition of plane waves. In the situation we need to replace complex numbers, e.g., $-i \omega$, $k$, $\varkappa$, and so on by nonlocal operators: $-i \omega \rightarrow \partial_t$, $k \rightarrow \hat k = (-\partial_x^2-\partial_y^2)^{1/2}$, $\varkappa \rightarrow \hat \varkappa = (\partial_t/\nu + \hat k^2)^{1/2}$, where the square roots should be taken with positive real parts. This implies that when an operator is acting on $h(t,x,y)$ it is necessary first to represent $h(t,x,y)$ as a superposition of plane waves, and then the action of the operator on each of the harmonics is determined by multiplying by the corresponding complex number. These notations are convenient and we will use them below.

\section{Nonlinear generation of the vertical vorticity}\label{sec:4}

It is natural to describe the horizontal eddy currents \cite{filatov2016nonlinear,filatov2016generation,francois2017wave} in terms of the vertical vorticity $\varpi_z$. However, in the linear approximation, according to expression (\ref{eq:linear_vorticity}), there is no contribution to this motion. In this section, we go beyond the approximation and take into account the main nonlinear contribution, which is of the second order in the wave steepness $|\nabla h| \ll 1$. As a result, we obtain the explicit formula for the vertical vorticity $\varpi_z$ in terms of the surface elevation $h(t,x,y)$ and the system parameters.

The vertical vorticity $\varpi_z$ must satisfy the exact equation (\ref{eq:vorticity_eq}). Then, up to the second order in the wave steepness $|\nabla h| \ll 1$, we obtain (see Appendix \ref{sec:appB})
\begin{equation}
 (\partial_z^2-\hat\varkappa^2)\varpi_z
 = - \nu^{-1} \varpi_\alpha \partial_\alpha v_z,
 \label{eq:nonlinear_vorticity}
\end{equation}
and we remind that the nonlocal operator $\hat\varkappa^2 = \partial_t/\nu + \hat k^2$. The term on the right-hand side is associated with the wave motion and can be regarded as a source with respect to the vertical vorticity $\varpi_z$, which is associated with the currents. The source can be interpreted as a rotation of two-dimensional vector $\varpi_{\alpha}$ by the velocity field of the surface waves. Roughly, one can say that the horizontal vorticity $\varpi_\alpha$, which is non-zero in the linear approximation and concentrated in thin viscous sublayer, is slightly rotated by the surface tilt and this gives rise to the vertical vorticity. The equation (\ref{eq:nonlinear_vorticity}) is the differential equation of the second order and thus it has to be supplemented by two boundary conditions. One of them was obtained earlier (\ref{boundary2}), and in the same order with respect to $|\nabla h| \ll 1$ we find the condition
 \begin{equation}
 \partial_z \varpi_z=
 \partial_\alpha h\partial_z \varpi_\alpha
 -\epsilon_{\alpha\gamma}
 (\partial_\alpha v_\beta+\partial_\beta v_\alpha)
 \partial_\beta \partial_\gamma h,
 \label{eq:boundary_vorticity}
 \end{equation}
which must be posed at $z=0$, see expression (\ref{B2}). The second condition is trivial: $\varpi_z \rightarrow 0$ at $z \rightarrow -\infty$. Note that the gradient of the film surface density drops from the boundary conditions (\ref{boundary2}) and (\ref{eq:boundary_vorticity}), and this means that the second order contribution in $|\nabla h| \ll 1$ to the film surface density $n(t,x,y)$ is irrelevant for the subsequent analysis.

The solution of the boundary value problem (\ref{eq:nonlinear_vorticity})-(\ref{eq:boundary_vorticity}) is straightforward. To calculate the right-hand sides one needs to know the velocity field and the vorticity in the linear approximation, which were obtained earlier, see expressions (\ref{eq:linear_velocity}) and (\ref{eq:linear_vorticity}). The details of the solution and the answer for the vertical vorticity $\varpi_z$, which is correct for an arbitrary form of the wave elevation $h(t,x,y)$, can be found in Appendix~\ref{sec:appB}. Here, motivated by recent experiments \cite{filatov2016nonlinear,filatov2016generation,francois2017wave}, we concentrate on a special form of the wave elevation $h(t,x,y)$, which allows to simplify the obtained expression (\ref{B7}) for the vertical vorticity $\varpi_z$. First, we assume that the pumping is monochromatic and then the frequencies of waves and their wave numbers are the same. Second, we believe that $\partial_x \partial_y h = 0$, e.g. the surface waves propagate perpendicular to each other. Then the expression for the vertical vorticity $\varpi_z$ takes a form
\begin{equation}\label{eq:Euler_Vort2}
  \varpi_z = \epsilon_{\alpha \beta} \left( e^{\hat\varkappa z} \frac{\hat\varkappa \hat D}{\hat k} \partial_{\beta} \partial_t h \right) \left( e^{\hat k z} \partial_{\alpha} h \right) + \epsilon_{\alpha \beta} \hat \varkappa^{-1} e^{\hat \varkappa z} \left( \left( \hat \varkappa \hat D \partial_{\beta} \partial_t h \right) \partial_{\alpha} h \right),
\end{equation}
and the relative accuracy of this expression is ${\mathcal O}(\gamma)$. Note that in the limiting cases of the free surface, $\hat D \rightarrow 2 \hat \varkappa k / (\hat \varkappa^2 + \hat k^2)$, and the almost incompressible film, $\hat D \rightarrow 1$, we can reproduce the previously obtained results \cite{filatov2016nonlinear,parfenyev2016effects}.

So far as we study the second-order nonlinearity, the vertical vorticity $\varpi_z$ varies in time with frequency $2 \omega$ or it is stationary. Depending on the characteristic frequency of the vorticity, the nonlocal operator $\hat \varkappa$ in the prefactor before brackets in the second term of expression (\ref{eq:Euler_Vort2}) must be estimated differently, because it acts on the second-order term in the wave elevation $h$. For the double-frequency contribution, $\varkappa \sim \sqrt{2\omega/\nu} \gg k$, and then the first term in expression (\ref{eq:Euler_Vort2}) is leading. For the stationary contribution, we can replace $\hat \varkappa$ by $\hat k$ in the prefactor before brackets in the second term, and then both terms in expression (\ref{eq:Euler_Vort2}) are of the same order. Furthermore, for the stationary contribution the first term is localized on the scale of the order of $\gamma/k$ near the surface in the viscous sublayer, while the second term penetrates deeper, on a distance of the order of $1/k$. Let us stress that the nonlinear stationary vorticity (\ref{eq:Euler_Vort2}) has a contribution, which penetrates in the bulk much deeper than the linear horizontal vorticity (\ref{eq:linear_vorticity}).

The value of the vertical vorticity on the fluid surface is especially important, because in most experiments the flows on the fluid surface are studied. It is given by expression (\ref{eq:Euler_Vort2}), where we must set $z=0$. Now it remains to understand how the nonlocal operators $\hat \varkappa$ and $\hat D$ act on the surface elevation $h(t,x,y)$. As earlier, we assume that the pumping is monochromatic, i.e the surface elevation $h(t,x,y)$ is given by a superposition of plane waves with frequency $\omega$ and wave number $k$, which are related to each other according to the dispersion law, $\omega^2 = gk + \sigma_0 k^3/\rho$. Then, by using the definition of operators, we obtain
\begin{eqnarray}
&\hat{\varkappa} h(t,x,y) =\displaystyle \frac{k}{\gamma} h \left( t+ \frac{\pi}{4 \omega}, x,y \right) + {\mathcal O}(\gamma kh),&\\
\label{eq:acting_D}
&\hat D h(t) =\displaystyle \frac{ \varepsilon^2 h(t) - \varepsilon h(t+\frac{\pi}{4 \omega}) - 2\gamma\varepsilon h(t - \frac{\pi}{2\omega}) + 2\gamma h(t- \frac{\pi}{4\omega}) }{\varepsilon^2-\varepsilon \sqrt{2} + 1} + {\mathcal O}(\gamma^2),&
\end{eqnarray}
and for brevity, we omitted the dependence on spatial coordinates in the second line.

These expressions allow to analyse how the film compressibility affects the generated vertical vorticity $\varpi_z$ on the fluid surface and in the bulk. However, in experiments the generated eddy currents are studied by observing the motion of particles immersed in the fluid. There is another second-order mechanism with respect to the wave steepness $|\nabla h| \ll 1$, which is well-known as the Stokes drift \cite{stokes1847theory}, that also influences the motion of particles. So, at first we need to investigate how the Stokes drift changes the eddy currents (see Sec.~\ref{sec:5}) and then we will discuss the resulting motion of particles, taking into account both contributions (see Sec.~\ref{sec:6}). Particular attention will be paid to the influence of the surface film on the vortex flow.

To conclude this section we would like to discuss the applicability conditions of the presented theory. On the one hand, we consider the limit of low viscosity, when the parameter $\gamma = \sqrt{\nu k^2/\omega} \ll 1$, but on the other --- we assume that the vortex flow is laminar. It means that the theory is correct if higher order nonlinear terms are small compared to the kept ones, i.e. the effective Reynolds number for the slow motion must be low. The characteristic velocity of the slow motion can be estimated from expression (\ref{eq:Euler_Vort2}) as $v^{(2)}~\sim~\varkappa D \omega h^2$ and therefore
\begin{equation}\label{eq:Reynolds}
\Re \sim \frac{({\bm v}^{(2)}\nabla)\varpi_z}{\nu\nabla^2 \varpi_z} \sim \frac{(kh)^2}{\gamma^3} D \ll 1.
\end{equation}
The dependence on the film properties is given by factor $D$. Note that the condition is always stronger than the small steepness condition $kh \ll 1$. If the condition (\ref{eq:Reynolds}) is violated then the vortices start to interact with each other and the motion becomes chaotic \cite{von2011double,francois2013inverse,francois2014three}.

\section{Mass transport}\label{sec:5}


In this section we study the vortex motion of passive tracers, which are immersed in the fluid with the surface elevation $h(t,x,y)$, up to the first two orders with respect to the wave steepness $|\nabla h| \ll 1$ and in the main approximation with respect to the parameter $\gamma \ll 1$, which means that the waves are weakly decaying. To characterize the position of a passive tracer we introduce a three-dimensional vector $\bm R$, which obeys the equation of motion
\begin{equation}\label{eq:Stokes}
\frac{d \bm R}{dt} = \bm v (t, \bm R).
\end{equation}
Near some point $\bm R(t_0) = \bm r_0$ we can expand the velocity field in the Taylor series
\begin{equation}
v_i (t, \bm r) = v_i (t, \bm r_0)+ G_{ij}(t,\bm r_0) (r_j - r_{0j}) + \dots,
\end{equation}
where $G_{ij}(t,\bm r_0) = \partial_j v_i (t, \bm r_0)$ is the velocity gradient tensor. Next, we solve equation (\ref{eq:Stokes}) up to the second order with respect to the parameter $|\nabla h| \ll 1$, using an iterative method. The displacement of the passive tracer is $\delta \bm R = \delta \bm R_0 + \delta \bm R_1$, where
\begin{equation}
\delta {\bm R}_{0} (t) = \int_{t_0}^t dt' \, \bm v (t', \bm r_0), \quad \delta R_{1i} (t) = \int_{t_0}^t dt' \, G_{ij} (t', \bm r_0) \delta R_{0j}(t'),
\end{equation}
and one needs to keep only linear terms in $\delta {\bm R}_{0}$ and $G_{ij}$ to calculate $\delta {\bm R}_1$. The velocity of the passive tracer, which initially was located in the point $\bm r_0$, is given by
\begin{equation}\label{eq:velocity_Lagrangian}
  \bm V_{L} (t) = \langle \bm v (t, \bm r_0) \rangle + \langle G_{ij} (t, \bm r_0) \delta R_{0j} (t) \rangle,
\end{equation}
where we perform averaging over fast wave oscillations denoted by angle brackets, since we are interested only in the slow motion. The last term in expression (\ref{eq:velocity_Lagrangian}) is known as the Stokes drift \cite{stokes1847theory}.

Experimentally the velocity field is reconstructed by analysing particle tracks. To capture only the slow motion it is convenient to synchronize recording camera with the wave period \cite{parfenyev2016effects}. In this case and if the used particles can be treated as passive tracers, the reconstructed velocity is described by expression (\ref{eq:velocity_Lagrangian}). As far as we would like to analyse horizontal eddy currents, we must calculate the vertical vorticity $\Omega_L$ of the measured velocity field, $\Omega_L = \epsilon_{\alpha \beta} \partial_{\alpha} V_{L\beta}$. Then the first term in expression (\ref{eq:velocity_Lagrangian}) produces the Eulerian vorticity, which was obtained in the previous section, see equation (\ref{eq:Euler_Vort2}), and the second term gives rise to the contribution $\varpi_S$ associated with the Stokes drift, i.e. $\Omega_L = \langle \varpi_z \rangle + \varpi_S$.

Next we proceed to the direct calculation of the vertical vorticity $\varpi_S = \epsilon_{\alpha \beta} \partial_{\alpha} \langle G_{\beta j}  \delta R_{0j}  \rangle$, which is produced by the Stokes drift. The details are presented in Appendix~\ref{sec:appC}. The reduction of the general result (\ref{St-dr-3}) to a specific case $\partial_{x}\partial_{y}h=0$ of e.g. orthogonal surface waves gives
\begin{equation}\label{eq:Stokes_vorticity}
  \varpi_S
  =
  \epsilon_{\alpha \beta}
  \left\langle \left( e^{\hat k z} \partial_{\beta} \partial_t h \right) \left( e^{\hat k z} \partial_{\alpha} h \right) \right\rangle
  -
  \epsilon_{\alpha \beta}
  \left\langle \left( e^{\hat\varkappa z} \frac{\hat\varkappa \hat D}{\hat k} \partial_{\beta} \partial_t h \right) \left( e^{\hat k z} \partial_{\alpha} h \right) \right\rangle,
\end{equation}
{and the relative accuracy of both terms is ${\mathcal O}(D\gamma)$. The first term is always leading outside the viscous sublayer, it is produced only by the potential part of the velocity field and penetrates on a depth of $1/k$. The second term is localized in the viscous sublayer on a depth of $\gamma/k$. Thus, the first term represents the well-known Stokes drift for the ideal fluid, while the second term takes into account the fluid viscosity and the presence of the surface film.}

Now we can calculate the vertical vorticity of the measured velocity field, $\Omega_L = \langle \varpi_z \rangle + \varpi_S$. By using relations (\ref{eq:Euler_Vort2}) and (\ref{eq:Stokes_vorticity}), we obtain
\begin{equation}\label{eq:Final}
  \Omega_L = \epsilon_{\alpha \beta} \left\langle \hat k^{-1} e^{\hat k z} \left( \left( \hat \varkappa \hat D \partial_{\beta} \partial_t h \right) \partial_{\alpha} h \right) \right\rangle + \epsilon_{\alpha \beta} \left\langle \left( e^{\hat k z} \partial_{\beta} \partial_t h \right) \left( e^{\hat k z} \partial_{\alpha} h \right) \right\rangle
  +{\mathcal O}(\omega (kh)^2D),
\end{equation}
where we have substituted $\hat \varkappa$ by $\hat k$ in the prefactor of the first term as we explained earlier. Both terms penetrate in the bulk on a distance of the order of $1/k$. Note that the terms in expressions (\ref{eq:Euler_Vort2}) and (\ref{eq:Stokes_vorticity}), which are localized in the viscous sublayer, have compensated each other and this reduction was known before for the case of fluid with a free surface \cite{longuet1953mass}.

In the case of the film with high compression modulus, $|D| \gg \gamma$, the first term in relation (\ref{eq:Final}) is leading and then the Stokes drift produces the negligible correction to the Eulerian vorticity. Note also that condition (\ref{eq:Reynolds}) of low Reynolds number can be rewritten in terms of the Lagrangian vorticity as $\Omega_L \ll \nu k^2$ for any value of the film compression modulus.


\section{Discussion}\label{sec:6}

In this section we illustrate the obtained results on particular examples and derive compact formulas, which can be used to describe experimental data. The section includes two parts. The first part is devoted to the linear analysis and here we discuss how the surface film alters the wave motion and how to infer the properties of thin liquid film if they are not known a-priori. In the second part we consider the intensity of eddy currents on the fluid surface and analyse how the eddy currents penetrate into the fluid bulk.

\subsection{Properties of the surface film and the wave motion}

The film on the fluid surface can be applied either specifically to control the transport of floating particles, or accidentally due to various contaminants and impurities. Various films may alter the hydrodynamic flow very differently depending on their properties. In the present paper, we consider thin (monomolecular) liquid films formed by insoluble agents, which can be characterized by two parameters: the equilibrium surface tension $\sigma_0$ and the dimensionless compression modulus $\varepsilon$.



Let us assume that the standing plane wave $h(t) = H \cos (kx) \cos (\omega t)$ is excited in the system. According to the performed analysis, the dispersion law of the surface waves is $\omega^2 = gk + \sigma_0 k^3/\rho$. Therefore, to infer an equilibrium value $\sigma_0$ of the surface tension coefficient, one needs to measure the wave frequency $\omega$ and the wave number $k$, and then substitute them into the dispersion law.

The value of parameter $\varepsilon$ can be also unknown. As it was shown in Sec.~\ref{sec:3}, the compression modulus of the film modifies the wave damping. By direct calculation, using expression (\ref{eq:wave_damping}), we find
\begin{equation}\label{eq:wave_damping2}
 \frac{\Im \ \omega}{\omega}
 = - 2\gamma^2 -\frac{\gamma}{2\sqrt{2}}\frac{\varepsilon^2}{(\varepsilon^2-\varepsilon\sqrt{2}+1)}
 +f(\varepsilon) \gamma^2 + {\mathcal O} (\gamma^3),
\end{equation}
where $f(\varepsilon)$ is some bounded analytic function and $f \rightarrow 0$ as $\varepsilon \rightarrow 0$. As compared to expression (\ref{eq:wave_damping}), we elaborated the order of the correction in accordance with expression (\ref{Dsmall}) at moderate values of film elastic stretching modulus, $\varepsilon \ll 1$, and for higher values of $\varepsilon$, the value of the operator $D$ was estimated by a number of the order of unity.

\begin{figure}
  \centerline{\includegraphics[width=0.95\linewidth]{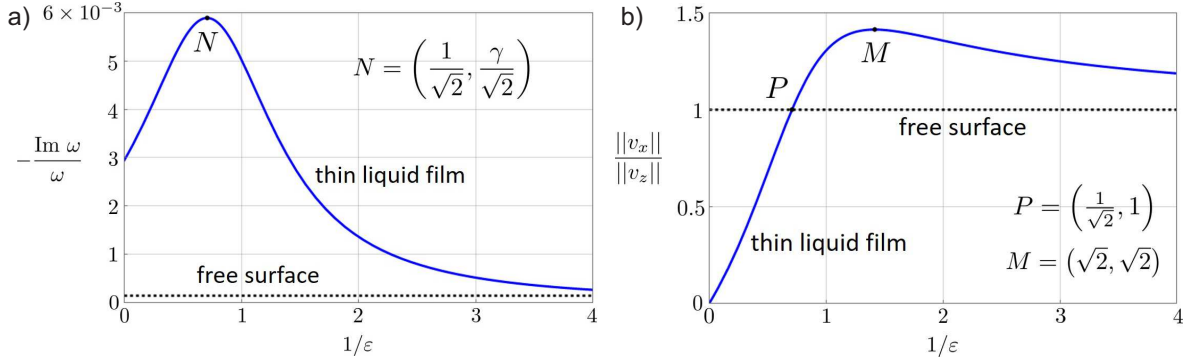}}
  \caption{a) The wave damping $-\Im \, \omega/\omega$ depending on the parameter $1/\varepsilon$, which characterises compressibility of the film. The parameter $\gamma$ is equal to $1/120$ that corresponds to the surface waves on water with frequency $3$ Hz. b) The ratio of amplitudes of horizontal and vertical velocities on the fluid surface depending on the same parameter. The black dashed lines correspond to the fluid with a free surface when $\varepsilon=0$.}
\label{fig:1}
\end{figure}

In equation (\ref{eq:wave_damping2}), the third term is always small as compared to the sum of the first and the second terms, its relative contribution is not larger than $\sqrt{\gamma}$ for any $\varepsilon$. The film considerably increases the wave decrement if $\varepsilon \gg \sqrt{\gamma}$, when the second term becomes the leading one. In this case, the decrement is shown in Fig.~\ref{fig:1}a. Note that the wave damping is maximized when $\varepsilon=\sqrt{2}$ under fixed values of the wave frequency $\omega$ and the wave number $k$. This means that the surface film with finite compressibility gives rise to stronger wave damping than the almost incompressible surface film, which corresponds to $\varepsilon\to\infty$. At $\varepsilon=0$ expression (\ref{eq:wave_damping2}) gives the answer for the free surface case. In general, equation (\ref{eq:wave_damping2}) has two different solutions with respect to the value of parameter $\varepsilon$. Therefore, sometimes in order to infer the film compressibility, it is not enough to measure experimentally the wave damping, the further analysis is required. Note also that expression (\ref{eq:wave_damping2}) does not take into account the dissipation related to the friction at the boundaries of the system, which can be important in laboratory experiments and depends on the size of the experimental setup.

According to expression (\ref{eq:linear_velocity}), the presence of the surface film strongly affects the horizontal velocity on the fluid surface. For example, in the case of the almost incompressible film, $D = 1$, one can obtain $v_{\alpha}|_{z=0} = 0$. Let us imagine that we have added floating passive tracers on the fluid surface. Then analysing the horizontal periodic motion with frequency $\omega$ of these tracers we can measure the horizontal velocity $v_{\alpha}$ on the fluid surface. The amplitude of horizontal velocity $||v_{\alpha}|| = \max_{x,y,t} |v_{\alpha}|$ will be proportional to the wave amplitude. Therefore to define the compressibility properties of the film we also should measure the wave elevation $h(t,x,y)$, which can be recalculated into the vertical velocity $v_z = \partial_t h$ on the fluid surface. By using expression (\ref{eq:linear_velocity}) and setting $z=0$, we can directly relate the ratio of horizontal and vertical velocities on the fluid surface and the parameter $\varepsilon$ of the surface film for the standing plane wave:
\begin{equation}\label{eq:ratio}
  \frac{||v_x||}{||v_z||}
  =\displaystyle \frac{1}{\sqrt{\varepsilon^2-\varepsilon \sqrt{2}+1}}(1 + {\mathcal O}(\gamma)).
\end{equation}
The velocities ratio (\ref{eq:ratio}) considerably differs from unity if $\varepsilon\gg1$, thus the higher value of compression modulus is needed to considerably change the ratio as compared to the value which leads to considerable increment in the wave damping, see expression (\ref{eq:wave_damping2}). Equation (\ref{eq:ratio}) allows to infer the value of parameter $\varepsilon$, which describes compressibility of the film, based on the experimentally measured ratio of velocity amplitudes $||v_x||/||v_z||$ of floating passive tracers. The ratio varies from $0$ to $\sqrt{2}$ depending on the film properties, see Fig.~\ref{fig:1}b. By combining two methods (\ref{eq:wave_damping2}) and (\ref{eq:ratio}), one can uniquely define the parameter $\varepsilon$ of the surface film.

\subsection{Eddy currents on the fluid surface and in the bulk}

\begin{figure}
  \centerline{\includegraphics[width=0.6\linewidth]{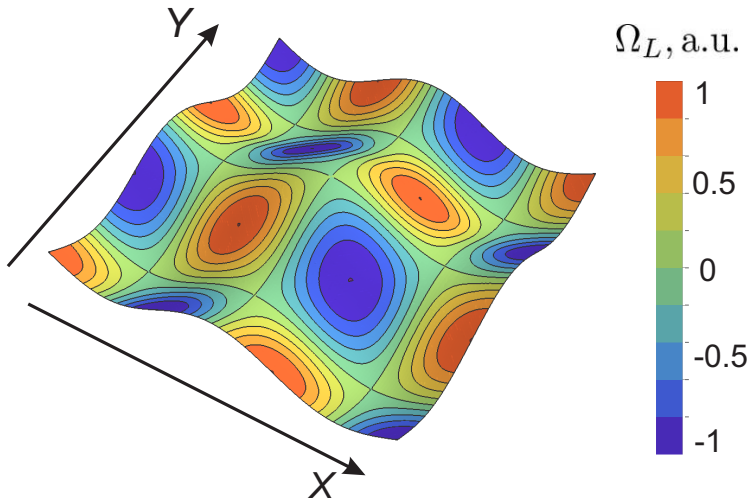}}
  \caption{Spatial structure of the generated eddy currents (\ref{eq:theory}), which are produced by two orthogonal standing waves (\ref{eq:orthogonal_waves}), propagating perpendicular to each other. The colors represent the intensity $\Omega_L$ of eddy currents in arbitrary units. Centers of vortices (where the vorticity is maximized) are located in the nodes of the wave pattern. Neighbouring vortices rotate in the opposite directions.}
\label{fig:2}
\end{figure}

Next we turn to the nonlinear analysis and here motivated by recent experiments \cite{filatov2016nonlinear,filatov2016generation,francois2017wave} we consider two monochromatic orthogonal standing waves propagating perpendicular to each other
\begin{equation}\label{eq:orthogonal_waves}
 h=H_1 \cos(\omega t) \cos(kx) + H_2 \cos(\omega t+\psi) \cos(ky).
\end{equation}
{One does not need to take into account viscous corrections to this expression corresponding to the wave spatial decay, which have relative magnitude of the order of ${\mathcal O}(D\gamma)$, see equation (\ref{eq:wave_damping}), since the corrections will produce parametrically smaller contribution to the generated eddy currents.} The phase shift $\psi$ is related to different boundary conditions for the standing waves in the $X$- and $Y$-directions. Substitution of equation (\ref{eq:orthogonal_waves}) into equation (\ref{eq:Final}) leads to expression for eddy currents that does not depend on time
\begin{equation}\label{eq:theory}
\Omega_L
= -
\left(\frac{\varepsilon^2 e^{kz\sqrt{2}}}{2 \gamma (\varepsilon^2-\varepsilon\sqrt{2}+1)} + \sqrt{2} e^{kz\sqrt{2}} + e^{2kz} \right) H_1 H_2 \omega k^2 \sin (kx) \sin (ky) \sin \psi
\end{equation}
and corresponds to a time-asymptotic value of the Lagrangian vorticity. 
The relative accuracy of equation (\ref{eq:theory}) is ${\mathcal O}(\sqrt{\gamma})$, the situation is similar to that for expression (\ref{eq:wave_damping2}), but here we omitted small terms. The sum of the first and the second terms correspond to the Eulerian vorticity, and the last term represents the contribution due to Stokes drift. Note that $Z-$dependence of these contributions is different. The surface film modifies the intensity of the eddy currents, leaving their $X-Y$ structure the same as compared to the free surface case \cite{filatov2016nonlinear}, see Fig.~\ref{fig:2}. If $\varepsilon \gg \sqrt{\gamma}$ then the first term in expression (\ref{eq:theory}) is leading. The limit $\varepsilon\gg1$ gives the answer for the case of fluid covered by the almost incompressible surface film \cite{parfenyev2016effects} and then the vorticity is proportional to $1/\sqrt{\nu}$. In the opposite case of almost free surface, $\varepsilon \ll \sqrt{\gamma}$, the first term in expression (\ref{eq:theory}) can be neglected and then the vorticity does not depend on the viscosity. To obtain the intensity of the eddy currents on the fluid surface one should set $z=0$ in expression (\ref{eq:theory}).

The dependence $\Omega_L \propto \sin \psi$ stated in equation (\ref{eq:theory}) immediately follows from the fact that the generation of eddy currents is a nonlinear second order effect, which was proved experimentally in papers \cite{filatov2016nonlinear,filatov2016generation,yablonskii2017acoustic}, and that the vortices' centers (where the vorticity is maximized) are located in the nodes of the wave pattern, which was also demonstrated experimentally in papers \cite{filatov2016nonlinear,filatov2016generation,francois2017wave}. Indeed, averaging over time of product of two monochromatic phase-shifted functions must be proportional to $\sin (\psi + \alpha)$, where $\psi$ is the phase shift and $\alpha$ is some possible correction, and therefore $\Omega_L \propto \sin (\psi + \alpha)$. Now, let us prove that for two orthogonal standing plane waves (\ref{eq:orthogonal_waves}) the parameter $\alpha$ is equal {to $\pi m, \, m \in \mathbb{Z}$}. Consider the symmetric situation, when the phase shift $\psi$ is equal to zero and when the amplitudes of surface waves are equal to each other. Then we can find any antinode of the wave pattern (like the center point in Fig.~\ref{fig:2}) and consider the rotation by an angle $\pi/2$ around the vertical axis passing through that point. With this transformation, the wave field does not change, but the eddy currents must change the sign. Therefore, the intensity of the eddy currents must be equal to zero and it proves that $\sin(\alpha) = 0$ and thus $\alpha = \pi m, \, m \in \mathbb{Z}$.

Let us discuss the correspondence between the experimental data and the dependence $\Omega_L \propto \sin \psi$. 
In the paper \cite{yablonskii2017acoustic}, where the authors studied the eddy currents in a thin smectic film performing transverse oscillations, it was reported that when the phase shift $\psi$ changes a sign then the eddy currents change their direction. In the other system \cite{filatov2016generation}, when the surface waves were excited on the deep water, the dependence $\Omega_L \propto \sin \psi$ was confirmed quantitatively. However, the obtained law (\ref{eq:theory}) cannot be applied to the experimental data directly, since the measurements were carried out after a time $t\sim 15\,\text{s}$ from the onset of excitation of the surface waves. The time is large as compared to the setting time of the wave pattern and thus the Stokes drift is determined by equation (\ref{eq:Stokes_vorticity}) as before. But the time is small as compared to the setting time of the Eulerian vorticity that is the viscous diffusion time, $t\ll 1/\nu k^2$. Indeed, the Eulerian vorticity is generated in the viscous sublayer according to equation (\ref{eq:nonlinear_vorticity}) and then it extends due to the viscosity in the fluid bulk. Thus, one should take into account that the waves were absent before the excitation, $h=0$ at $t<0$ in expression (\ref{eq:orthogonal_waves}), so the Eulerian part of the vorticity spread only to a depth $d \sim \sqrt{\nu t}$ out of the viscous sublayer at the time of measurements, see equation (\ref{eq:nonlinear_vorticity}). Hence, the Reynolds number (\ref{eq:Reynolds}) should be now evaluated as $\mathrm{Re}\sim \Omega_L d^2/\nu \sim \Omega_L t$ (we have used the fact that the nonlinear velocity field is two-dimensional) and it remains of the order or less than unity according to the experimental data. Therefore, the measurements were carried out in weakly-nonlinear regime and our general arguments based on the symmetry consideration remain valid and they explains the experimental data. Note also, that the absolute value of the measured vorticity cannot be directly compared with expression (\ref{eq:theory}), since the Eulerian contribution is suppressed in comparison with its time-asymptotic value.

In a different study of the same system \cite{francois2017wave}, the authors observed that the surface flow becomes disordered and {contains a large-scale contribution of size $\gg 1/k$} but remains stationary, when the phase shift $|\psi| < 50^\circ$. It was reported that the dependence of the vortex-grid amplitude on the phase shift $\psi$ does not correspond to expression (\ref{eq:theory}) for these angles. We speculate that the large-scale flow may be related to an imperfection of the excitation process since the symmetry of the flow does not correspond to the symmetry of two orthogonal standing waves (\ref{eq:orthogonal_waves}). According to the experimental data, when the phase shift becomes sufficiently small, these additional large-scale currents increase in amplitude and deform the predicted regular small-scale pattern. Indeed, the large-scale velocity dominates at small $\psi$ and thus its typical amplitude is $\sqrt{E}$, where $E$ is the mean velocity squared. Therefore the advection time by the large-scale velocity is $1/k \sqrt{E}\sim 2 \,\text{s}$ at a scale $1/k$, and it is small as compared to the vortex-grid viscous setting time $1/\nu k^2\sim 150\,\text{s}$. This means that the nonlinear interaction between the large-scale and small-scale velocity components of the mass transport is essential and the result (\ref{eq:theory}) obtained in weakly-nonlinear regime is not applicable.

Next, let us discuss the amplitude of the generated eddy currents. Comparing expressions (\ref{eq:theory}) and (\ref{eq:wave_damping2}) one can see that the intensity of the eddy currents is closely related to the damping of surface waves. Both these quantities increase parametrically when $\varepsilon \gg \sqrt{\gamma}$ in comparison with the free surface case \cite{filatov2016nonlinear}. The surface film leads to the additional tangential stresses, which parametrically increase the non-potential contribution to the velocity field and the horizontal vorticity, and therefore also increase the dissipation of energy. Now the main dissipation of energy occurs in the viscous sublayer. The vertical vorticity is generated due to the rotation of horizontal vorticity, and for this reason it also increases. Note that the intensity of eddy currents have a maximum when $\varepsilon=\sqrt{2}$. The physics behind this maximum is the same as in the case of amplification of the damping of surface waves.

The experimental data for capillary waves \cite{filatov2016nonlinear} shows considerable excess of the Lagrangian vorticity on the fluid surface as compared to the theoretical result for the free surface case. Indeed, using this data one obtains that the factor in the round brackets in expression (\ref{eq:theory}) is equal to about $13$, while the theory leads to the factor $1+\sqrt{2}$ for $\varepsilon=0$ and $z=0$. In the experiment, the stationary regime was examined and the Reynolds number (\ref{eq:Reynolds}) was less or of the order of unity. The similar discrepancy can be observed for gravitational waves \cite{filatov2016generation}, although the measurements were made in a transient non-stationary regime.
Based on the developed theory, we would like to speculate that the discrepancy can be attributed to the influence of a film presented on the fluid surface, but further experimental studies are required. The other possible reason for the discrepancy is the influence of capillary effects on the motion of floating particles \cite{falkovich2005surface}. The latter effect could be eliminated if the measurements are carried out in the fluid bulk. Note that for the discussed experiment \cite{filatov2016nonlinear}, the model of the almost incompressible film yields the factor of about $10.5$. However, the surface film of finite elasticity can produce more intense eddy currents. The maximum value of the factor is approximately $18.7$ and it corresponds to $\varepsilon = \sqrt{2}$.

To conclude we would like to add that the Stokes drift alone calculated for ideal fluid (produced by the potential part of velocity) explains the generation of eddy currents qualitatively since it leads to the same spatial structure of the measured vertical vorticity \cite{francois2017wave}. This contribution was missed in \cite{filatov2016nonlinear} and first it was pointed out by Oliver Buhler in private communication. The consistent account of viscosity shows that it is significant in the calculation of Stokes drift since it produces correction inside the viscous sublayer, see expression (\ref{eq:Stokes_vorticity}) and paper \cite{parfenyev2016effects}. In the free surface case this correction changes the sign of the Stokes drift contribution on the fluid surface; in the contaminated case, when the fluid surface is covered by a thin liquid film with high compression modulus, $\varepsilon \gg \sqrt{\gamma}$, the viscous correction is leading. Nevertheless, the viscous correction to the Stokes drift seems to be unphysical, because it is completely compensated by an analogous correction into the Eulerian vorticity, see expression (\ref{eq:Final}). As a result, the total Lagrangian vorticity is a sum of contributions from Eulerian motion and Stokes drift, which both penetrate on a depth of the order of $1/k$ and have the same sign. Also, we would like to stress that the contribution associated with the Eulerian vertical vorticity, see the first term in expression (\ref{eq:Final}), is crucial to explain the phenomenon quantitatively. If the surface is covered by a thin film with a high compression modulus, $\varepsilon \gg \sqrt{\gamma}$, then one can neglect the second term in expression (\ref{eq:Final}) and the Eulerian vorticity is leading, while in the case of the almost free surface, $\varepsilon \ll \sqrt{\gamma}$, both contributions are of the same order.

\section{Conclusion}\label{sec:7}

To summarise we studied how a thin {insoluble} film {with zero shear elasticity and small shear and dilational viscosities} presented on a fluid surface modifies the wave motion and affects the nonlinear generation of eddy currents. We showed that the presence of the film significantly changes the velocity field in a viscous sublayer (\ref{eq:linear_velocity}) and increases the damping of the surface waves (\ref{eq:wave_damping2}) in comparison with the free surface case, see Fig.~\ref{fig:1}a. We also found that the ratio of amplitudes of horizontal and vertical velocities on the fluid surface (\ref{eq:ratio}) can take any value from $0$ to $\sqrt{2}$ depending on the film properties, see Fig.~\ref{fig:1}b. Based on these observations we propose two experimental methods to infer the film properties by analysing the wave damping and the motion of floating passive tracers. Let us stress that all the results are valid for both capillary and gravitational waves.

We also generalized the theory of vertical vorticity generation \cite{filatov2016nonlinear}, taking into account compressibility of the film. We obtained the explicit expression (\ref{eq:Euler_Vort2}) for the vertical vorticity $\varpi_z$ in terms of the surface elevation $h(t,x,y)$ and established the applicability condition (\ref{eq:Reynolds}) of our approach. We also analysed the motion of passive tracers in the fluid bulk and calculated the vortical contribution to the motion of these tracers associated with the Stokes drift (\ref{eq:Stokes_vorticity}). We analysed the expressions and obtained the compact formula (\ref{eq:theory}) suitable for describing the experimental data. The results form a quantitative basis for the analysis of eddy currents, which are generated on the fluid surface and in the bulk due to nonlinear interaction of waves. The resent experimental progress in the field \cite{filatov2016nonlinear,filatov2016generation,francois2017wave} encourages us to think that test of our theoretical findings is the matter of near future.

\acknowledgments
We thank V. Lebedev and S. Filatov for valuable discussions. This work was supported by the Russian Science Foundation, Grant No. 14-22-00259. V. Parfenyev acknowledges support from the Foundation for the advancement of theoretical physics and mathematics ''BASIS''.

\appendix
\section{Gravity-capillary surface waves}\label{sec:appA}

In this section we linearize the system of equations (\ref{eq:Navier-Stokes})-(\ref{eq:mass_conservation}) and find a solution in the form of a plane wave $\propto \exp(i k_{\alpha} r_{\alpha} - i \omega t)$ corresponding to the gravity-capillary surface wave, where $\omega$ is the wave frequency and $k=(k_x^2+k_y^2)^{1/2}$ is its wavenumber. The linearized Navier-Stokes equation (\ref{eq:Navier-Stokes}) has a form
\begin{equation}\label{eq:Navier-Stokes2}
  \partial_t \bm v = - \nabla P/\rho + \nu \nabla^2 \bm v.
\end{equation}
By taking the divergence and using the incompressibility condition $\mathrm{div} \, \bm v = 0$, we find that the pressure $P$ should be a solution of the Laplace equation $\nabla^2 P = 0$,
\begin{equation}\label{eq:solution}
P = P_0 e^{k z} e^{i k_\alpha r_\alpha - i \omega t}, \quad z \leq 0.
\end{equation}
Substituting the solution (\ref{eq:solution}) into the linearized Navier-Stokes equation (\ref{eq:Navier-Stokes2}), we find
\begin{eqnarray}
   (\partial_t + \nu k^2 - \nu \partial_z^2) \tilde{v}_{\alpha} &=& -ik_{\alpha} P_{0} e^{kz}/\rho, \\
   (\partial_t + \nu k^2 - \nu \partial_z^2) \tilde{v}_{z} &=& -k P_{0} e^{kz}/\rho,
\end{eqnarray}
where $v_i = \tilde{v}_i \exp(i k_{\alpha} r_{\alpha} - i \omega t)$.
The system of equations has a solution, which is a sum of forced (potential) and eigen (solenoidal) terms
\begin{equation}\label{eq:A5}
   \tilde{v}_{\alpha} = \frac{k_{\alpha} P_{0}}{\rho \omega} e^{kz} + \varkappa \tilde{A}_{\alpha} e^{\varkappa z}, \quad
   \tilde{v}_{z} = \frac{-i k P_{0}}{\rho \omega} e^{kz} - i k_{\alpha} \tilde{A}_{\alpha} e^{\varkappa z},
\end{equation}
where we have used the incompressibility condition $i k_{\alpha} \tilde{v}_{\alpha} + \partial_z \tilde{v}_z = 0$, and we have also introduced $\varkappa = \sqrt{k^2 - i \omega/\nu}$ (with positive real part). To find the values of constants $\tilde{A}_{\alpha}$ we should use the boundary conditions.

In the linear approximation the boundary conditions should be posed at $z=0$. Let us denote the equilibrium values of the film surface density and the surface tension as $n_0$ and $\sigma_0$ correspondingly. Then $n=n_0 + \delta n$, $\sigma = \sigma_0 + \sigma'(n_0) \delta n$, and the boundary conditions (\ref{eq:DynBC_t}) and (\ref{eq:mass_conservation}) give
\begin{eqnarray}
  \label{eq:A6}
  \rho \nu (\partial_{\alpha} v_z + \partial_z v_{\alpha}) &=& \sigma'(n_0) \partial_{\alpha} \delta n,\\
  \label{eq:A7}
  \partial_t \delta n + n_0 \partial_{\alpha} v_{\alpha} &=& 0.
\end{eqnarray}
Next, by substituting solution (\ref{eq:A5}) into equations (\ref{eq:A6}) and (\ref{eq:A7}), one obtains
\begin{equation}\label{eq:A8}
  \delta \tilde{n} = \frac{n_0}{\omega} \left( \frac{k^2 P_0}{\rho \omega} + \varkappa k_{\alpha} \tilde{A}_{\alpha} \right), \quad \tilde{A}_{\alpha} = -\frac{k_{\alpha} P_0}{\rho \omega \varkappa} D,
\end{equation}
where $\delta n = \delta \tilde{n} \exp(i k_{\alpha} r_{\alpha} - i \omega t)$ and parameter $D$ depends on the compressibility properties of the surface film
\begin{equation}
  D = \frac{2i\gamma - \varepsilon}{i\gamma \frac{\varkappa^2 + k^2}{\varkappa k}-\varepsilon},
\quad
\varepsilon = \frac{-n_0 \sigma'(n_0)}{\rho \sqrt{\nu \omega^3}/k^2}.
\end{equation}

The relation between the pressure $P_0$ and the surface elevation $h$ can be obtained from the kinematic boundary condition (\ref{eq:kinematic_BC}). In the linear approximation it takes the form $\partial_t h = v_z$ and should be posed at $z=0$. Finally, we obtain
\begin{equation}\label{eq:A10}
  P_0 e^{i k_\alpha r_\alpha - i \omega t} = \frac{i \rho \omega \partial_t h}{k \left(1 - \frac{k}{\varkappa} D \right)}.
\end{equation}
By combining relations (\ref{eq:A5}), (\ref{eq:A8}) and (\ref{eq:A10}), we can rewrite the expression for the velocity field in terms of the surface elevation $h$:
\begin{equation}
v_{\alpha} = \frac{\left( e^{{k}z} - {D} e^{{\varkappa} z} \right)}{{k} \left(1 - \frac{ k}{\varkappa} {D} \right)} \partial_{\alpha} \partial_t h, \quad
v_z = \frac{\left( e^{{k}z} - \frac{ k}{\varkappa} {D} e^{{\varkappa} z} \right)}{\left(1 - \frac{ k}{ \varkappa} {D} \right)} \partial_t h.
\end{equation}

The last boundary condition (\ref{eq:DynBC_n}) allows one to obtain the dispersion law of the gravity-capillary waves. In the linear approximation it takes a form
\begin{equation}
P - 2 \rho \nu \partial_z v_z = \rho g h - \sigma_0 \nabla^2 h,
\end{equation}
and then one can find
\begin{equation}\label{eq:Al}
    \omega^2
    \ = \
    \left(1-\frac{k}{\varkappa}D\right)\left(gk +\frac{\sigma_0}{\rho}k^3\right)
    -2i(1-D)\nu k^2\omega.
\end{equation}
Let us remind that we are interested in the behaviour of weakly decaying waves, i.e. the parameter $\gamma = \sqrt{\nu k^2/\omega} \ll 1$ and therefore $|k/\varkappa| \sim \gamma \ll 1$. Moreover, the parameter $\varepsilon$ is real because we have assumed that there is no time lag between changes in $n$ and $\sigma(n)$. Thus, the absolute value of the parameter $D$ is of the order or less than unity. Then, in the main approximation with respect to the parameter $\gamma \ll 1$, we obtain
\begin{equation}
\omega^2 = gk +(\sigma_0/\rho) k^3.
\end{equation}
One can also find the small imaginary part of the wave frequency $\omega$, which describes the wave damping. By using expression (\ref{eq:Al}) and approximate relation $k/\varkappa \approx \gamma e^{i \pi/4}$, one finds
\begin{equation}
\label{eq:A15}
 \frac{\Im \ \omega}{\omega}
 = - \frac{\gamma}{2 \sqrt 2} \left( \Re D + \Im D \right)
 -
 \gamma^2 + {\mathcal O}(D\gamma^2),
\end{equation}
where only leading terms must be kept. In particular, this means that if $\Re D + \Im D \sim 1$ then it is incorrect to keep the second term in the expression, since terms of the same order were omitted in deriving (\ref{eq:A15}). On the other hand, if $\Re D + \Im D \sim \gamma$ then the first two terms in equation (\ref{eq:A15}) are comparable to each other and the omitted correction is small in comparison with any of them. Note that the condition $\gamma \ll 1$ indeed means that the waves are weakly decaying.

\section{Solution of the vorticity equation}\label{sec:appB}

The vertical vorticity $\varpi_z$ is generated due to nonlinear interaction of surface waves in the viscous fluid. By using $z$-component of equation (\ref{eq:vorticity_eq}) and keeping only the second order terms with respect to the wave steepness, one obtains
\begin{equation}\label{Stokes-with-force}
    (\partial_t - \nu \nabla^2) \varpi_z = (\bm\varpi \cdot \nabla) v_z,
\end{equation}
where we take into account that $\varpi_z$ is zero in linear approximation, see expression (\ref{eq:linear_vorticity}).
Using the notations, which were introduced in Sec.~\ref{sec:3}, one can rewrite the equation in the following form
\begin{equation}\label{B1}
 (\partial_z^2-\hat\varkappa^2)\varpi_z
 = - f, \quad f = \nu^{-1} \varpi_\alpha \partial_\alpha v_z.
\end{equation}
This equation should be supplemented by the boundary condition (\ref{boundary2}). In the linear approximation the components of the unit vector normal to the surface are $l_\alpha = -\partial_\alpha h$, $l_z=1$, and the curvature tensor has nonzero components $K_{\alpha\beta}=-\partial_\alpha\partial_\beta h$. Keeping all second order terms in the wave steepness, we obtain the following boundary condition
\begin{equation}\label{B2}
 \left(\partial_z \varpi_z - \partial_\alpha h\partial_z \varpi_\alpha\right)\big\vert_{z=0}
 +
 \epsilon_{\alpha\gamma}
 (\partial_\alpha v_\beta+\partial_\beta v_\alpha)
 \partial_\beta \partial_\gamma h \big\vert_{z=0}=
 0, \quad \varpi_z (-\infty) = 0.
\end{equation}
The solution of equation (\ref{B1}) is $\varpi_z (z)= e^{\hat{\varkappa} z} A(z) + e^{-\hat{\varkappa} z} B(z)$, where
\begin{equation}\label{B3}
\partial_z A = - \hat{\varkappa}^{-1} e^{-\hat \varkappa z} (f/2), \quad \partial_z B = \hat{\varkappa}^{-1} e^{\hat \varkappa z} (f/2).
\end{equation}
By using expressions (\ref{eq:linear_velocity}) and (\ref{eq:linear_vorticity}) for the velocity field and the horizontal vorticity in the linear approximation, up to the first two orders in the parameter $\gamma \ll 1$, we obtain:
\begin{equation}
f = \nu^{-1} \epsilon_{\alpha \beta} \left( \left[\frac{\hat\varkappa \hat D}{\hat k} + \hat D^2 \right] e^{\hat\varkappa z} \partial_{\beta} \partial_t h \right) \left( \left[ e^{\hat k z} + \frac{\hat k}{\hat\varkappa}\hat D(e^{\hat k z}-e^{\hat\varkappa z})\right]\partial_{\alpha} \partial_t h \right).
\end{equation}
The second terms in the square brackets should be kept only if the estimate $|D| \gg \gamma$ is valid. Integrating relations (\ref{B3}), we find under the same approximation in the parameter $\gamma \ll 1$:
\begin{equation}\label{B5}
  \varpi_z(z) = e^{\hat \varkappa z}F + \epsilon_{\alpha \beta} \Bigg[ \frac{\hat k_2 \hat D_2}{2 \hat k_1} + 
  \frac{\hat\varkappa_1}{\hat k_1}\left(1 + \frac{\hat k_2}{\hat \varkappa_2} \hat D_2 +\frac{\hat k_1}{\hat \varkappa_1} \hat D_1-
  2 \frac{\hat k_2}{\hat\varkappa_1} \right) \Bigg] \hat D_1 e^{(\hat\varkappa_1+\hat\varkappa_2)z} \partial_{\beta} \partial_t h \partial_{\alpha} h,
\end{equation}
where operators with subscript '$1$' or '$2$' act only on the first or the second multiplier $h$ respectively. We kept first-order corrections with respect to the parameter $\gamma$ in expression (\ref{B5}), since they are necessary to find the function $F$ from the boundary condition (\ref{B2}), because the leading terms in the parameter $\gamma$ are cancelled. The boundary condition in terms of surface elevation $h$ with the same accuracy reads
\begin{equation}\label{B6}
 \partial_z \varpi_z (0) = \epsilon_{\alpha \beta} \left( \frac{\hat\varkappa_1^2 \hat D_1}{\hat k_1} + \hat\varkappa_1 \hat D_1^2 \right) \partial_{\beta} \partial_t h \partial_{\alpha} h - 2 \epsilon_{\alpha \beta} \frac{(1- \hat D_1)}{\hat k_1} \partial_{\alpha} \partial_\gamma \partial_t h \partial_{\beta} \partial_\gamma h.
\end{equation}
Substituting relation (\ref{B5}) into boundary condition (\ref{B6}), we finally obtain:
\begin{eqnarray}
\label{B7}
\nonumber
 \varpi_z &=& \epsilon_{\alpha \beta} \left( e^{\hat\varkappa z} \frac{\hat\varkappa \hat D}{\hat k} \partial_{\beta} \partial_t h \right) \left( e^{\hat k z} \partial_{\alpha} h \right) + 2 \epsilon_{\alpha \beta} \hat \varkappa^{-1} e^{\hat \varkappa z} (1-\hat D_2) \partial_{\alpha} \partial_{\gamma} h \, \partial_{\beta} \partial_{\gamma} \partial_t \hat{k}^{-1} h \\ && + \epsilon_{\alpha \beta} \hat \varkappa^{-1} e^{\hat \varkappa z} \Big(\hat D_1 \hat \varkappa_1 \frac{\hat k_2}{\hat k_1} - \hat D_1 \hat D_2 \frac{\hat k_2}{2 \hat k_1} (\hat \varkappa_1- \hat \varkappa_2) \Big) \partial_{\beta} \partial_t h \partial_{\alpha} h.
 \end{eqnarray}
The relative accuracy of the expression is ${\mathcal O}(\gamma)$ and thus only leading terms must be kept (they are different depending on the film properties and the characteristic frequency of the vertical vorticity). Let us stress that result (\ref{B7}) is correct for an arbitrary form of the wave elevation $h(t,x,y)$. In the particular case of the monochromatic pumping, the frequencies of waves and their wave numbers are the same, and then the last term proportional to $\hat\varkappa_1 - \hat\varkappa_2$ is equal to zero.

\section{Stokes drift calculation}\label{sec:appC}

We conduct direct calculation of Stokes drift in this section. As earlier we assume that the pumping is monochromatic.  By using expression (\ref{eq:linear_velocity}) for the velocity field we find the displacement of a passive tracer in the linear approximation with respect to the wave amplitude and in the leading order with respect to the parameter $\gamma \ll 1$
\begin{equation}\label{St-dr-1}
\delta {\bm R}_0
=
\left(\frac{(e^{\hat k z}-\hat D e^{\hat\varkappa z})}{\hat k} \partial_{x} h,
\frac{(e^{\hat k z}-\hat D e^{\hat\varkappa z})}{\hat k} \partial_{y} h,
e^{\hat k z} h \right)^{T} + \mathrm{const},
\end{equation}
where we do not take into account the constant term, since it produces zero correction after averaging over time in equation (\ref{eq:velocity_Lagrangian}). The velocity gradient tensor $G_{ij}(t,\bm r_0) = \partial_j v_i (t, \bm r_0)$ under the same approximation is given by
\begin{equation}
G_{ij} = \left(
\begin{array}{ccc}
 \frac{e^{\hat k z}- \hat D e^{\hat\varkappa z}}{\hat k} \partial_{xx}^2 \partial_t h & \frac{e^{\hat k z}- \hat D e^{\hat\varkappa z}}{\hat k} \partial_{xy}^2 \partial_t h & \left(e^{\hat k z} - \frac{\hat \varkappa \hat D}{ \hat k} e^{\hat\varkappa z} \right) \partial_x \partial_t h \\
 \frac{e^{\hat k z}- \hat D e^{\hat\varkappa z}}{\hat k} \partial_{xy}^2 \partial_t h & \frac{e^{\hat k z}- \hat D e^{\hat\varkappa z}}{\hat k} \partial_{yy}^2 \partial_t h & \left( e^{\hat k z} - \frac{\hat\varkappa \hat D}{\hat k} e^{\hat\varkappa z} \right) \partial_y \partial_t h \\
 * & * & *
\end{array} \right),
\end{equation}
and then the vertical vorticity $\varpi_S = \epsilon_{\alpha \beta} \partial_{\alpha} \langle G_{\beta j}  \delta R_{0j}  \rangle$, which is produced by the Stokes drift, is equal to
\begin{eqnarray}\label{St-dr-3}
\nonumber
    \varpi_S
    & = &
    \epsilon_{\alpha \beta} \left\langle \left( e^{\hat k z} \partial_{\beta} \partial_t h \right) \left( e^{\hat k z} \partial_{\alpha} h \right) \right\rangle
    -
    \epsilon_{\alpha \beta}
    \left\langle
    \left( e^{\hat\varkappa z} \frac{\hat\varkappa \hat D}{\hat k} \partial_{\beta} \partial_t h \right) \left( e^{\hat k z} \partial_{\alpha} h \right)
    \right\rangle
    \\&&
    +
    \epsilon_{\alpha \beta}
    \left\langle
    \left( \frac{(e^{\hat k z}-\hat D e^{\hat\varkappa z})}{\hat k} \partial_{\beta}\partial_\gamma \partial_t h \right)
    \left( \frac{(e^{\hat k z}-\hat D e^{\hat\varkappa z})}{\hat k} \partial_\alpha \partial_\gamma h \right)
    \right\rangle.
\end{eqnarray}
First term in expression (\ref{St-dr-3}) describes the contribution produced by the wave motion of ideal fluid, when $\partial_x \partial_y h=0$, e.g., by two waves propagating perpendicular to each other. The second term represents the correction, which arises due to the fluid viscosity and the film compressibility. The last term appears if the excited surface waves are not orthogonal.


%

\end{document}